\documentclass[orivec,runningheads]{llncs}
\usepackage[T1]{fontenc}
\usepackage{graphicx}
\usepackage{booktabs}
\usepackage[misc]{ifsym}
\newcommand{\corr}{(\Letter)}
\usepackage{mwe}

\usepackage{soul}
\usepackage{url}
\usepackage[hidelinks]{hyperref}
\usepackage{amssymb}
\usepackage{booktabs}
\usepackage{algorithm}
\usepackage{algorithmic}
\usepackage{multirow}
\usepackage{multicol}
\usepackage{float}
\usepackage{balance}

\usepackage{mathtools}
\newcommand{\rcn}{\texttt{BunCa}}

\newcommand\blfootnote[1]{%
  \begingroup
  \renewcommand\thefootnote{}\footnote{#1}%
  \addtocounter{footnote}{-1}%
  \endgroup
}

\begin{document}

\title{Bundle Recommendation with Item-level Causation-enhanced Multi-view Learning}

\titlerunning{Causation-enhanced Multi-view Learning}


\author{Huy-Son Nguyen$^*$  \and 
Tuan-Nghia Bui$^*$ \and
Long-Hai Nguyen \and
Hung Hoang \and \\
Cam-Van Thi Nguyen \and
Hoang-Quynh Le \and
Duc-Trong Le \corr }
\institute{University of Engineering and Technology, \\ 
Hanoi Vietnam National University, Vietnam \\
\email{\{huyson,21020364,21020624,21020518,vanntc,lhquynh,trongld\}@vnu.edu.vn}}



\authorrunning{H.S. Nguyen et al.}

\tocauthor{Huy-Son Nguyen, Tuan-Nghia Bui, Long-Hai Nguyen, Hung Hoang, Cam-Van Thi Nguyen, Hoang-Quynh Le, Duc-Trong Le} 
\toctitle{Bundle Recommendation with Item-level Causation-enhanced Multi-view Learning} 

\maketitle              
\blfootnote{$^\ast$ Equal Contribution.}

\begin{abstract}
Bundle recommendation aims to enhance business profitability and user convenience by suggesting a set of interconnected items.
In real-world scenarios, leveraging the impact of asymmetric item affiliations is crucial for effective bundle modeling and understanding user preferences. To address this, we present \rcn{}, a novel bundle recommendation approach employing item-level causation-enhanced multi-view learning. \rcn{} provides comprehensive representations of users and bundles through two views: the Coherent View, leveraging the Multi-Prospect Causation Network for causation-sensitive relations among items, and the Cohesive View, employing LightGCN for information propagation among users and bundles. 
Modeling user preferences and bundle construction combined from both views ensures rigorous cohesion in direct user-bundle interactions through the Cohesive View and captures explicit intents through the Coherent View.
Simultaneously, the integration of concrete and discrete contrastive learning optimizes the consistency and self-discrimination of multi-view representations. Extensive experiments with \rcn{} on three benchmark datasets demonstrate the effectiveness of this novel research and validate our hypothesis.
\end{abstract}

\section{Introduction}
\label{sec:intro}
Recommendation systems are crucial in enhancing user experiences and shaping business strategies, particularly in e-commerce \cite{zhu2014bundle,sun2022revisiting}. 
While conventional recommendation systems traditionally prioritize individual item suggestions, bundle recommendations have been recognized as a superior strategic marketing tactic.
Bundle recommendation, rooted in user behavior and item relevance, involves grouping pertinent items into bundles, such as assortments of items from the same category (e.g., \textit{detective books, thrilling games}) \cite{chen2019matching,sun2022revisiting}, complementary item bundles (e.g., \textit{men's vest with cravat, phone with case}) \cite{sun2022revisiting}, etc. 
Recommending bundles based on user preferences poses greater challenges compared to separate items due to data sparsity and the diverse composition of bundles.

\begin{figure}[h]
	\centering
	\includegraphics[width=0.8\textwidth]{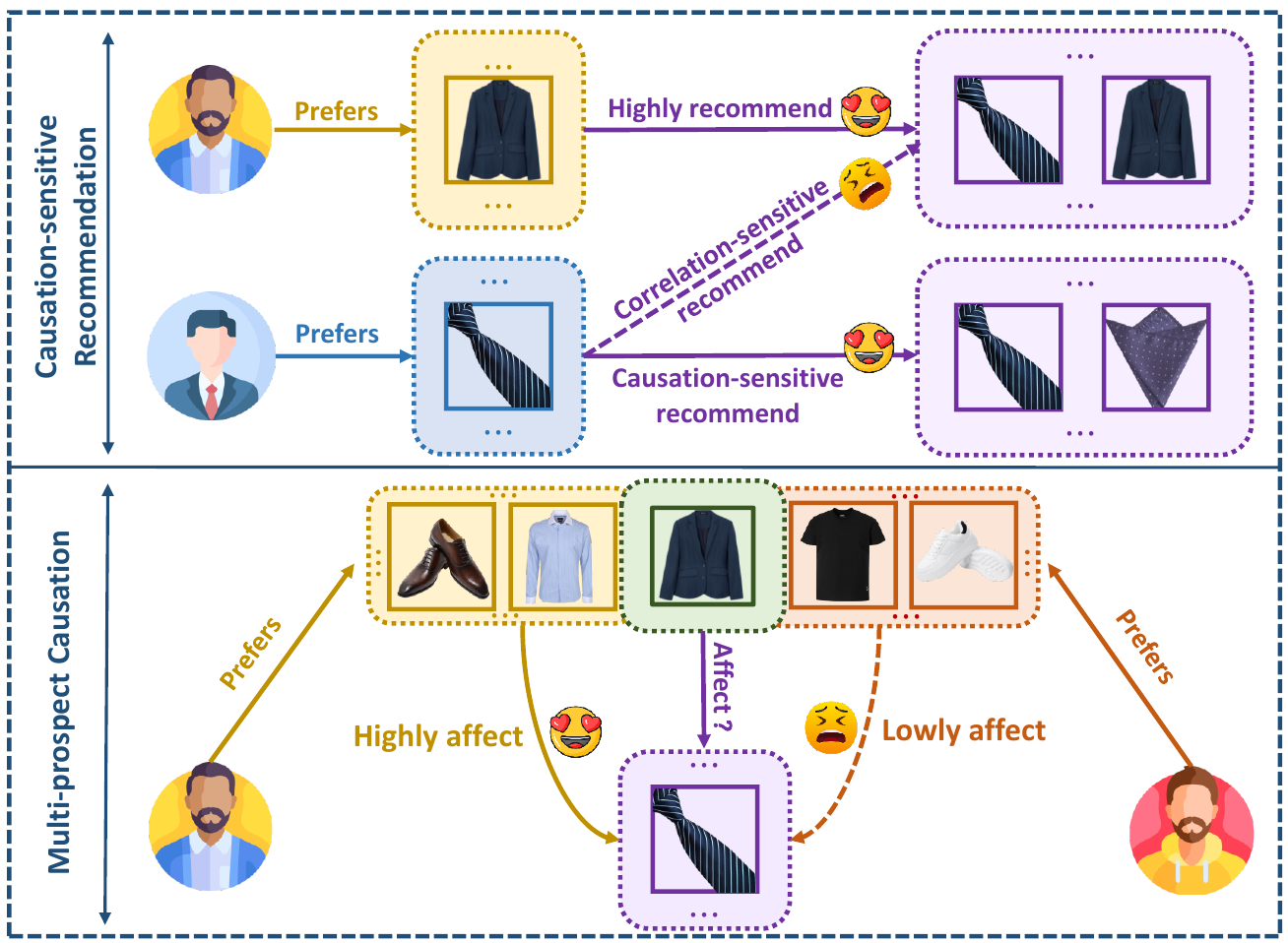}
	\caption{Motivating examples of multi-prospect causation in bundle recommendation. \label{fig:motivating_example}}
\end{figure}

Prior studies have mainly focused on selecting items for bundling \cite{bai2019personalized} and identifying product bundles in commercial markets \cite{tzaban2020product}. However, recent works \cite{chang2020bundle,ma2022crosscbr} have shifted towards recommending existing bundles, aiming to better align suggested bundles with users' preferences.
State-of-the-art (SOTA) bundle recommendation such as MIDGN \cite{zhao2022multi}, CrossCBR \cite{ma2022crosscbr}, and BundleGT \cite{wei2023strategy} aggregate items to generate user and bundle representations. However, they often overlook the intricate relationships between individual items within bundles. 
There has been an initial exploration of the impact of item relationships, often grounded in symmetric correlations, on purchasing decisions for a bundle \cite{le2019correlation,ariannezhad2023complex}.
In real-world scenarios, the influence between items within bundles is usually asymmetric, with anchor items playing a significant role. 
Symmetric correlation-based recommendation models, as illustrated in Figure~\ref{fig:motivating_example}, fall short in accurately capturing user preferences.
While \textit{blazers} and \textit{cravats} may be frequently purchased together, their influence on each other is not necessarily symmetrical. 
Users with an interest to \textit{blazers} might consider to combine with \textit{cravats} as accessories, while others primarily interested in \textit{cravats} may already have suitable \textit{blazers}, making the combination with \textit{pocket squares} more logical. 
Understanding these dynamics requires examining item-item relations across diverse user preferences, which may help improve the bundle recommendation performance.

\textbf{Approach and Contributions}. Motivated by mentioned limitations of previous works, we introduce \rcn{} a novel neural architecture for bundle recommendation, whereby leverages asymmetric relationships between bundle items via the item-level causation-enhanced multi-view learning. 
To the best of our knowledge, this paper represents the first comprehensive study to explore asymmetric item relationships, which is \textit{our first contribution}.
In Section \ref{sec:method}, these information is transmitted in a multi-view learning framework which consists of two distinct views namely Coherent View and Cohesive View. 
The Cohesive View learning aims to encode rigorously high-order collaborative signals in user-bundle interactions.
Besides, the Multi-Prospect Causation Networks for causation-sensitive relations between items are integrated to generate enhanced twin representations of objects in the Coherent View learning. 
Modeling user preferences and bundle construction combined from both views ensures rigorous cohesion in direct interactions between users-bundles via the Cohesive View, as well as explicit meaning of their intents via the Coherent View.
\textit{As our second contribution}, we employ concrete and discrete contrastive learning simultaneously, enhancing the consistency and self-discrimination of representations from both views. Section \ref{sec-exp} presents \textit{our final contribution} in conducting extensive experiments on three benchmark datasets. The experimental results demonstrate that \rcn{} has potential comparison with state-of-the-art methods for the bundle recommendation task, substantiating the rationalization of our proposed approach and new prominent issues in existing bundle-related tasks.

\section{Related Work}
\label{sec-rw}


The field of bundle recommendation has witnessed diverse approaches aimed at accurately recommending pre-defined item sets to users.
Early approaches \cite{liu2014recommending}, grounded in the BPR framework \cite{rendle2012bpr}, deliberate users' past interactions with lists of items and individual items.
DAM model \cite{chen2019matching} emerged as a pioneer by recognizing the importance of affiliated items within bundles and optimizing both user-item and user-bundle interactions using attention mechanisms and multi-task learning. 
Recently, BGCN \cite{chang2020bundle} used Graph Convolutional Network operations to capture intricate relations between users, items, and bundles from two distinct views, establishing an effective multi-view learning approach for subsequent SOTA models.

In the realm of multi-view architectures, addressing challenges related to the inconsistency and integration of information concerning objects has been tackled through the adoption of contrastive learning (CL) techniques.
CrossCBR \cite{ma2022crosscbr}, evolving from BGCN, stands out for its remarkable improvements, leveraging cross-view contrastive learning to highlight the significance of capturing cooperative information from two distinct views. 
MIDGN \cite{zhao2022multi} introduces the concept of intent disentanglement, modeling multiple latent features for each view and employing contrastive loss between the two views. 
Our study introduces concrete contrastive learning to underscore the self-discrimination of augmented representations in multi-view learning. 
In addition, discrete contrastive learning is employed to address the problem of inconsistency between separate views.


The capture of intricate connections between items influencing bundling strategies and user preferences has gained attention.
MIDGN~\cite{zhao2022multi}, which tackles hidden intents inside user preference and bundle construction, encounters performance and complexity challenges.
BundleGT~\cite{wei2023strategy} achieves SOTA performance with a Hierarchical Graph Transformer network. 
Regarding the recommendation of item sets, Beacon \cite{le2019correlation} utilizes correlation-sensitive signals to enhance the prediction of user with next-basket. 
It is noteworthy that relationships between items in real-world scenarios are not symmetric, in contrast to the simulation in \cite{le2019correlation,wei2023strategy}.
Integrating homogeneous graphs is also employed to leverage item-item relationships, solely utilizing the LightGCN operation \cite{zhou2023enhancing,nguyen2023hhmc}.
Furthermore, some further approaches such as EBRec augmenting observed user-item interactions with pre-trained models \cite{du2023enhancing}, or DGMAE disillation framework \cite{ren2023distillation} also introduce several new aspects.
Our study focuses on exploiting asymmetric item-item relationships via the features of user preference and bundle construction, leading to enhanced user/bundle representations compared to previous studies.

\section{Methodology}
\label{sec:method}

\begin{figure*}[h]
	\centering
	\includegraphics[width=1.0\columnwidth]{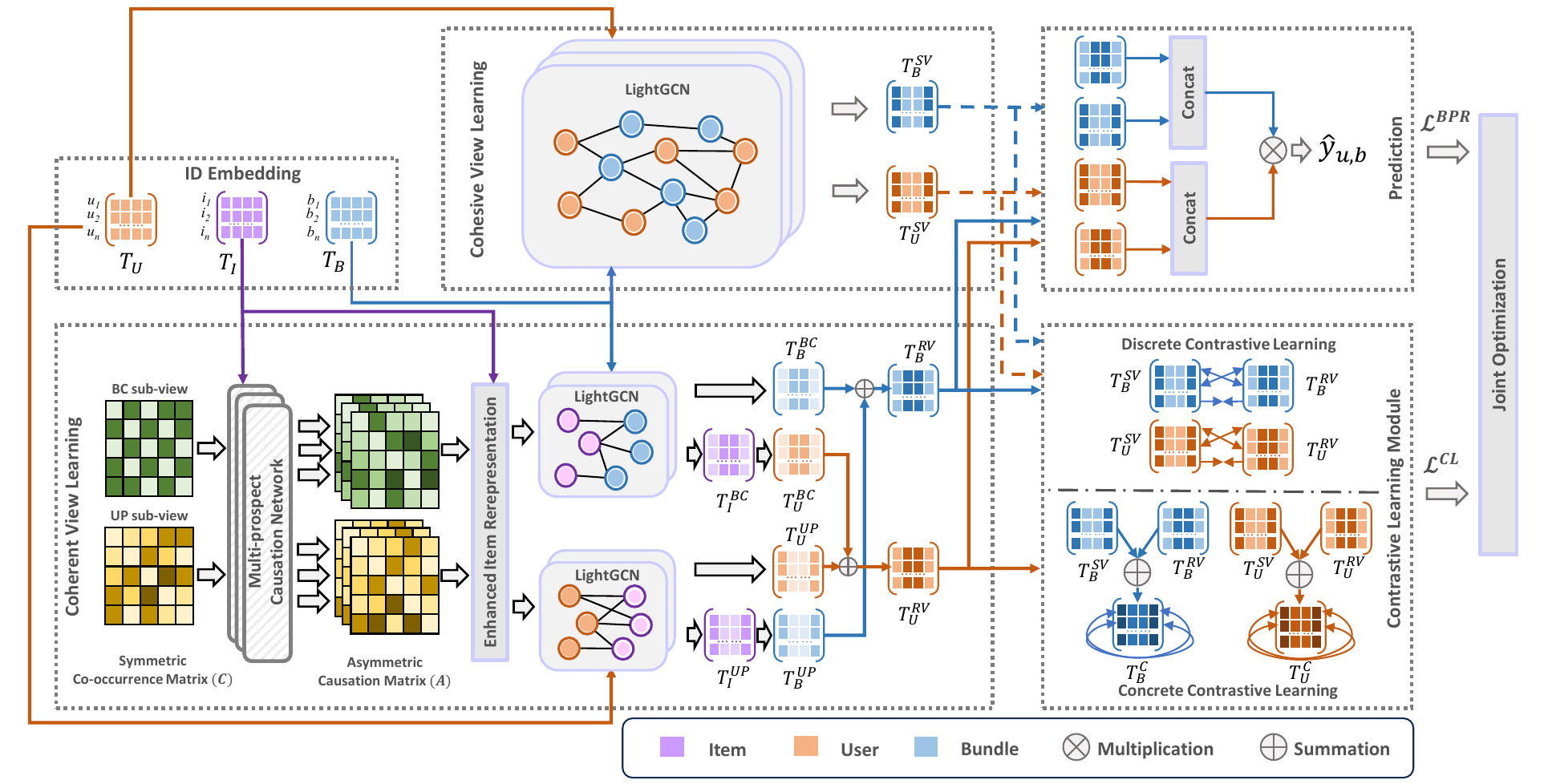}
    \caption{The schematic illustration of our proposed model \rcn{}.} 
    \label{fig:model}
\end{figure*}

In this section, we initially formalize the bundle recommendation problem and define the detailed interaction graph construction.
Subsequently, we demonstrate the four important modules of \rcn{} illustrated in Figure~\ref{fig:model} including: (1) Cohesive View representation learning; (2) Coherent View representation learning; (3) contrastive learning modules; and (4) prediction and joint optimization module. 



\subsection{Preliminaries}
\label{ssec:pre}


\subsubsection{Problem Formulation.}
Given the set of users  $\mathcal{U} = \{{u_1}, {u_2}, \ldots, {u_{|\mathcal{U}|}}\}$,  the set of bundles $\mathcal{B} = \{{b_1}, {b_2},\ldots, {b_{|\mathcal{B}|}}\}$, and the set of items $\mathcal{I} = \{{i_1}, {i_2}, \ldots, {i_{|\mathcal{I}|}}\}$.
The user-bundle interactions, user-item interactions, and bundle-item affiliations are respectively defined as three binary-valued matrices $ X \in \{0,1\}^{|\mathcal{U}| \times |\mathcal{B}|}, Y \in \{0,1\}^{|\mathcal{U}|\times|\mathcal{I}|}$, and $Z \in \{0,1\}^{|\mathcal{B}|\times|\mathcal{I}|}$
where the cell with value $1$ denotes an observed relation between the user-bundle, user-item, or bundle-item pair, value $0$ for otherwise. The objective of our work is to accurately predict unseen user-bundle interactions for recommendation system.


\subsubsection{Graph Construction in the Cohesive View.} Inspired by collaborative filtering concepts suggesting that users with mutual interactions share similar preferences \cite{wang2019neural}, we assume that user preference patterns are hidden within the co-occurrence of purchased bundles. 
In addition, bundles often bought collectively by numerous individuals may indicate shared interests. 
To capture these hypotheses, the user co-occurrence matrix $C_{\mathcal{U}} = X \cdot X^\intercal \in \mathbb{R}^{|\mathcal{U}\times \mathcal{|U|}}$ and the bundle co-occurrence matrix $C_{\mathcal{B}} = X^\intercal \cdot X \in \mathbb{R}^{|\mathcal{B}\times \mathcal{|B|}}$ are both derived from user-bundle interactions. 
Thereby, we explicitly establishes sets of connections denoted as: $\mathcal{E}_{\mathcal{UB}}=\{e_{u,b} | b \in \mathcal{B} \land u \in \mathcal{U} \land X_{(u,b)} = 1)\}$, $\mathcal{E}_{\mathcal{U}}=\{e_{u,u'} | (u, u' \in \mathcal{U}) \land C_{\mathcal{U}(u,u')} \geq 1)\}$, $\mathcal{E}_{\mathcal{B}}=\{e_{b,b'} | (b, b' \in \mathcal{B}) \land C_{\mathcal{B}(b,b')} \geq 1)\}$. 
\rcn{} integrates user-bundle interactions with homogeneous correlations to thoroughly exploit the intrinsic relationships among users and bundles in graph $\bar{\mathcal{G}}=\{ \bar{\mathcal{V}}, \bar{\mathcal{E}} \}$, where $ \bar{\mathcal{V}}= \{\mathcal{U} \cup \mathcal{B}\}$ and $\bar{\mathcal{E}} = \{ \mathcal{E}_{\mathcal{UB}} \cup  \mathcal{E}_{\mathcal{U}} \cup \mathcal{E}_{\mathcal{B}} \}$ respectively describe the set of vertices and the set of edges.

\subsubsection{Graph Construction in the Coherent View.} The bundle-item bipartite graph $\mathcal{G}_{\mathcal{B}\mathcal{I}}$, and user-item bipartite graph $\mathcal{G}_{\mathcal{U}\mathcal{I}}$ are leveraged to effectively aggregate information on item-level representations.
The graph $\mathcal{G}_{\mathcal{BI}}=\{\mathcal{V}_{\mathcal{BI}}, \mathcal{E}_{\mathcal{BI}}\}$ is constructed by the affiliations between bundles and items in $Z$, where $\mathcal{V}_{\mathcal{BI}}=\{\mathcal{B} \cup \mathcal{I}\}$ and $\mathcal{E}_{\mathcal{BI}} = \{e_{b,i} | b \in \mathcal{B}, i \in \ \mathcal{I} \}$  represent the set of vertices and edges.
User preferences, derived from historical interactions with standalone items, also impact decision-making for bundle purchases. Thus, the graph  $\mathcal{G}_{\mathcal{UI}} = \{ \mathcal{V}_{\mathcal{UI}}, \mathcal{E}_{\mathcal{UI}} \}$, where $\mathcal{V}_{\mathcal{UI}} = \{\mathcal{U} \cup \mathcal{I}$\} and $\mathcal{E}_{\mathcal{UI}} = \{e_{u,i} | u \in \mathcal{U}, i \in \ \mathcal{I} \}$ are established based on the interactions between users and items in $Y$. 
Furthermore, at the beginning of the training phase, the features of users $T_{\mathcal{U}} \in \mathbb{R}^{|\mathcal{U}|\times d}$, bundles $T_{\mathcal{B}} \in \mathbb{R}^{|\mathcal{B}|\times d}$, and items $T_{\mathcal{I}} \in \mathbb{R}^{|\mathcal{I}|\times d}$ are randomly initialized, where $d$ is the embedding dimensionality.


\subsection{Cohesive View Representation Learning}
\label{ssec:cohesive}

As mentioned, the heterogeneous graph $\bar{\mathcal{G}}$ may help enhance robust propagation of high-order collaborative signals by incorporating homogeneous correlations. \rcn{} adopts the LightGCN operation \cite{he2020lightgcn} to encode cohesive representations of users and bundles, excluding self-connections and non-linear transformations in the propagation function.
The $h$-th layer’s information propagation of $\bar{\mathcal{G}}$ is:
\begin{equation}
    t^{SV(h)}_v = \sum_{v' \in \mathcal{N}_{v}} \frac{t_{v'}^{SV(h-1)}}{\sqrt{|\mathcal{N}_{v}|} \sqrt{|\mathcal{N}_{v'}|}}
\label{eq:1}
\end{equation}
where $t^{SV(h)}_v \in \mathbb{R}^{d}$, $t^{SV(h)}_{v'} \in \mathbb{R}^{d}$ represent the $h$-th layer’s information propagated to nodes $v, v' \in \bar{\mathcal{V}}$; 
${\mathcal{N}}_{v}$ and $\mathcal{N}_{v'}$ denote the first-hop neighbors of node $v$ and $v'$ in $\bar{\mathcal{G}}$ respectively.
Thereby, the embedding of user $t^{SV}_u$ and bundle $t^{SV}_b$ learned via Cohesive View are inferred as follows:
\begin{equation}
    t^{SV}_u = \sum_{h=0}^H t^{SV(h)}_u \newline~~,~~
    t^{SV}_b = \sum_{h=0}^H t^{SV(h)}_b
\label{eq:ub_final_SV}
\end{equation}
where, the initial embeddings $t^{SV(0)}_u$ and $t^{SV(0)}_b$ are obtained from $T_{\mathcal{U}}$ and $T_{\mathcal{B}}$.

Overall, the representations of all users $T^{SV}_{\mathcal{U}} \in \mathbb{R}^{|\mathcal{U}| \times d}$ and bundles $T^{SV}_{\mathcal{B}} \in \mathbb{R}^{|\mathcal{B}| \times d}$ through the unification of all embeddings of different layers rigorously integrate information propagated from high-order neighbors.

\subsection{Coherent View Representation Learning}
\label{ssec:coherent}
For the Coherent View, \rcn{} aims to explicitly leverage the asymmetric relation between items based on the user preference (UP) sub-view and the bundle construction (BC) sub-view to enhance the modeling of decision-making factors.
Learning these sub-views improves item embeddings, hence consolidating the representations for users and bundles. These representations are fused to highlight vital information that determines user preferences and bundle construction.
Initially, we employ a coarse-grained mixture method to create item-item symmetric relations. The co-occurrence item matrices from different aspects are computed by the matrix multiplication as follows:
\begin{equation}
    C_{\mathcal{I}}^{UP} = Y^\intercal \cdot Y \newline~~,~~
    C_{\mathcal{I}}^{BC} = Z^\intercal \cdot Z
\end{equation}
where $C_{\mathcal{I}}^{UP}, C_{\mathcal{I}}^{BC} \in \mathbb{R}^{|\mathcal{I}| \times |\mathcal{I}|}$ represent item co-occurrence based on user-item interactions $Y$ and bundle-item affiliations $Z$.
The symmetric connections between items are normalized as follows:

\begin{equation}
    \Tilde{C}_{(i,j)}= 
    \begin{cases}
        1 & \text{if ${C}_{(i,j)} \geq 1 \land i \ne j$ },  \\ 
        0 & \text{otherwise} 
        \end{cases}
\end{equation}  
where $i,j \in \mathcal{I}$ refer to the corresponding items. 
Filtering the threshold of ${C}_{(i,j)}$ can be considered empirically to achieve better performance on different data domains.
It is notable that the symmetric matrix $\Tilde{C}_{\mathcal{I}}$ is uniform symbol for $C_{\mathcal{I}}^{UP}, C_{\mathcal{I}}^{BC}$ due to the similar calculations of two sub-views in many scenarios.

\subsubsection{Multi-Prospect Causation Network.}

Assuming that causation-sensitive relationships exist among items frequently purchased together, \rcn{} employs Multi-Prospect Causation Network (MPCNet) to explicitly model asymmetric associations between items.
MPCNet is constructed with $L$ prospects, each represented by a learnable prospect vector $\mathbf{p}_l \in \mathbb{R}^{d}$. 
For the $l$-th prospect, the weight $r_{j \to i}$ signifies the influence from item $j$ to item $i$ based on various user preferences and bundling strategies, derived as follows:

\begin{equation}
    r^l_{j\to i} = \mathbf{p}^{\intercal}_l \cdot \varphi(\Psi^l_{src} \cdot t_j \oplus \Psi^l_{dst}\cdot t_i \oplus \Phi)
\end{equation}
where $\Psi^l_{src}, \Psi^l_{dst} \in \mathbb{R}^{d \times d}$ are learnable parameters for the source and destination object in the $l$-th prospect; 
$t_i, t_j \in \mathbb{R}^d$ denote the representations of item $i$ and item $j$; $\varphi(.)$ is the non-linear activation function; $\oplus$ denotes the element-wise summation; and $\Phi$ is the bias parameter.

In the $l$-th prospect, the asymmetric causation matrix ${A}^{l} \in \mathbb{R}^{|\mathcal{I}| \times |\mathcal{I}|}$, representing the causation-sensitive relationships among items at fine-grained level, is computed by the attention mechanism concept of GATv2 \cite{brody2021attentive}. The weight ${A}^{l}_{(i,j)}$ describes how much item $i$ is influenced by item $j$, defined as follows:
\begin{equation}
    {A}^{l}_{(i,j)} = \frac{exp(r^l_{j\to i}) \odot \Tilde{C}_{(i,j)}  }{max(\sum_{j' \in \mathcal{I}} exp(r^l_{j' \to i}) \odot \Tilde{C}_{(i,j')}, \epsilon)}
\end{equation}
where $\epsilon \in R$ is a fixed constant; $\odot$ display the multiplication between two scalars.

\subsubsection{Enhancing Item Representation.}
The asymmetric relationships obtained from MPC-Net are utilized to encode the latent representation of item $t_i^{l} \in \mathbb{R}^d$ in the $l$-th prospect, formulated as follows:
\begin{equation}
t_i^{l} = \sum_{j \in \mathcal{I}} A^{l}_{(i, j)} \Psi^{l}_{src} \cdot t_j
\end{equation}
where $\Psi^l_{src} \in \mathbb{R}^{d \times d}$ denotes the dense layer to learn item features; and $t_j \in \mathbb{R}^d$ denotes other item representation.
Subsequently, the multi-prospect item representation is devised using the residual connection method as:
\begin{equation}
    t^{f}_i = \alpha \frac{1}{L} \sum_{k=1}^{L} t_i^{l} \oplus (1-\alpha) t_i
\end{equation}
where $t^f_i \in \mathbb{R}^d$ is the enhanced item representation, combining individual features and aggregated information from connected items with a controlled influence $\alpha \in [0, 1]$.
It is notable that ${t}^f_i$ serves as a uniform placeholder, representing both $\Hat{t}^{UP}_i \in \Hat{T}^{UP}_{\mathcal{I}}$ and $\Hat{t}^{BC}_i \in \Hat{T}^{BC}_ \mathcal{I}$, denoting the enhanced item representation of the UP sub-view and BC sub-view. 

For UP sub-view, the item representation $\Hat{t}_i^{UP}$ is employed alongside the user's preference $t_u \in T_{\mathcal{U}}$ as the node's initial feature within the user-item graph $\mathcal{G}_{\mathcal{UI}}$. 
The graph-based propagation process utilizes the LightGCN operation similar to Eq~(\ref{eq:1}).
The ultimate representation of users and items is denoted as ${T}^{UP}_{\mathcal{U}} \in  \mathbb{R}^{|\mathcal{U}|\times d}$ and 
${T}^{UP}_{\mathcal{I}} \in  \mathbb{R}^{|\mathcal{I}|\times d}$.
The item features is aggregated to derive the bundle features $T_{\mathcal{B}}^{UP} \in \mathbb{R}^{|\mathcal{B}|\times d}$ utilizing the mean pooling operation as:
\begin{equation}
    t_b^{UP} = \sum_{i \in \mathcal{N}_b^{BI}}  \frac{t^{UP}_i}{|\mathcal{N}_b^{BI}|}
\end{equation}
where $t_b^{UP},t_i^{UP} \in \mathbb{R}^d$ describe the aggregated representation of bundle $b$ and item $i$ in the UP sub-view; $\mathcal{N}_b^{BI}$ is the neighbor set of bundle $b$ in graph $\mathcal{G}_{\mathcal{BI}}$.

Similar operations are employed in the BC sub-view learning procedure. 
The graph $\mathcal{G}_{\mathcal{BI}}$ takes $t_b \in T_{\mathcal{B}}$ and $\Hat{t}_i^{BC}$ as input representations for nodes in the first layer. 
The information propagation is performed using the LightGCN operation, which results in the final representations of bundles and items respectively denoted as ${T}^{BC}_{\mathcal{B}} \in  \mathbb{R}^{|\mathcal{B}|\times d}$ and ${T}^{BC}_{\mathcal{I}} \in  \mathbb{R}^{|\mathcal{I}|\times d}$.
Subsequently, the final user representations $T_{\mathcal{U}}^{BC} \in \mathbb{R}^{|\mathcal{U}|\times d}$ in the BC sub-view is derived as follows:
\begin{equation}
    t_u^{BC} = \sum_{i \in \mathcal{N}_u^{UI}}  \frac{t^{BC}_i}{|\mathcal{N}_u^{UI}|}
\end{equation}
where $t_u^{BC} \in T_{\mathcal{U}}^{BC}, t_i^{BC} \in {T}^{BC}_{\mathcal{I}}$ describe the aggregated representation of user $u$ and item $i$ in BC sub-view; $\mathcal{N}_u^{UI}$ denotes the neighbors of user $u$ in graph $\mathcal{G}_{\mathcal{UI}}$.

The comprehensive information is integrated by the attentive summation as:
\begin{equation}
\label{eq:attsum}
\begin{split}
    t^{RV}_u = \beta t^{BC}_u \oplus (1-\beta) t^{UP}_u, &\\ 
    t^{RV}_b = \beta t^{BC}_b \oplus (1-\beta) t^{UP}_b & 
\end{split}
\end{equation}
where the hyper-parameter $\beta \in [0,1]$ is to maintain a balance between the propagated features from the two sub-view learning phases. $t^{RV}_u , t^{RV}_b \in \mathbb{R}^d$ denotes the final representation of user $u$ and bundle $b$ through Coherent View learning.


\subsection{Contrastive Learning Module}

The discrete contrastive learning is directed at minimizing the inconsistency between two different views, whereas the concrete contrastive learning strives to enhance the distinctiveness of augmented representations.

\subsubsection{Discrete Contrastive Learning.}
Inspired by CrossCBR \cite{ma2022crosscbr}, \rcn{} addresses variations in user preferences for items and bundles by minimizing inconsistencies in representations across different views. 
The in-batch negative sampling is adopted to construct the negative pairs \cite{wu2021self}. 
Relied on InfoNCE \cite{gutmann2010noise}, the discrete contrastive loss function for user representations is computed from contrastive pairs ($t^{SV}_u, t^{RV}_u$) as follows:
\begin{equation}
\begin{split}
    \mathcal{L}_{u}^{DC} = -\frac{1}{|\mathcal{U}|} \sum_{{u} \in \mathcal{U}} log \frac{exp((\cos(t^{SV}_{u} , t^{RV}_{u} )/\tau))}{\sum_{{u}' \in \mathcal{U}} exp((\cos(t^{SV}_{u} , t^{RV}_{{u}'})/ \tau))}
\end{split}
\end{equation}
where ${u}'$ represents a user other than ${u}$; $\cos(., .)$ denotes the cosine similarity function; $\tau$ is the temperature hyper-parameter that help to distinguish the positive and negative samples. 
The formula for calculating $\mathcal{L}_{b}^{DC}$ is derived similarly.

\subsubsection{Concrete Contrastive Learning}
The concrete contrastive learning is devised to distinguish the unique characteristics of each individual user and bundle. 
We create the combination of user preferences and bundle features as follows:
\begin{equation}
    t_u^C = t^{RV}_u \oplus t_u^{SV} ~~\newline,~~
    t_b^C = t^{RV}_b \oplus t^{SV}_b
\end{equation}
where $t_u^C , t_b^C$ denote the fused multi-view representation of user and bundle, respectively.
The formulation of the concrete contrastive loss function for user representations is derived as follows:
\begin{equation}
\begin{split}
    \mathcal{L}_{u}^{CC} = -\frac{1}{|\mathcal{U}|} \sum_{{u} \in \mathcal{U}} log \frac{exp(1/\tau)}{\sum_{{{u}'} \in \mathcal{U}} exp((\cos(t^C_{{u}}, t^C_{{u}'})/ \tau))} &\\
\end{split}
\end{equation}
where ${u}'$ denotes a different user of ${u}$; \( \cos(., .) \) denotes the cosine similarity function; and \( \tau \) is the temperature hyper-parameter that aids in distinguishing between positive and negative samples.
The same operation is repeated for the concrete contrastive loss of bundle $\mathcal{L}_{b}^{CC}$.

The final contrastive loss is derived from the dyadic contrastive learning techniques described above as follows:
\begin{equation}
    \mathcal{L}^{CL} = \gamma (\frac{\mathcal{L}^{DC}_u + \mathcal{L}^{DC}_b}{2})
    +(1-\gamma) (\frac{\mathcal{L}^{CC}_u +\mathcal{L}^{CC}_b}{2})
\end{equation}
where $\gamma \in [0,1]$ controls which contrastive learning is more important.

\subsection{Prediction and Joint Optimization}

For the comprehensive representations, we effectively capture augmented information from both views, derived as follows:
\begin{equation}
\begin{split}
    \Tilde{t}_u = [~\mu t^{SV}_u ~\mathbin\Vert~ t^{RV}_u~] \newline~~,~~
    \Tilde{t}_b = [~\mu t^{SV}_b ~\mathbin\Vert~ t^{RV}_b~]
\end{split}
\end{equation}
where the hyper-parameter $\mu\in [0,1]$ controls the impact of views on the final user's preference and bundle's feature; $\mathbin\Vert$ denotes the concatenation; and $\Tilde{t}_u , \Tilde{t}_b \in \mathbb{R}^{2d}$ respectively represent the fusion of user's preferences and bundle's features, combining Cohesive View and Coherent View learning.

Finally, the interaction probability $\mathbf{\hat{y}}_{u,b} \in \mathbb{R}$ between user $u$ and bundle $b$ is calculated by the inner-product as follows:
\begin{equation}
    \mathbf{\hat{y}}_{u,b} = (\Tilde{t}_u)^\intercal \cdot \Tilde{t}_b
\end{equation}



The Bayesian Personalized Ranking loss \cite{rendle2012bpr} is utilized to enhance probabilities for interacted user-bundle pairs while reducing for non-interacted pairs:
\begin{equation}
    \mathcal{L}^{BPR} = \sum_{(u, b, b') \in \mathcal{S}} -ln \sigma(\mathbf{\hat{y}}_{u,b} - \mathbf{\hat{y}}_{u, b'})
\end{equation}
where $\mathcal{S} = \{(u,b ,b') | u\in \mathcal{U}; b, b' \in \mathcal{B}; (u,b) \in X; (u,b') \notin X \}$ and $\sigma(.)$  is the $sigmoid$ activation function.
The final loss $\mathcal{L}$ is combined by weighted summation of $\mathcal{L}^{BPR}, \mathcal{L}^{CL}$ and the $L2$ regularization $||\Theta||^2_2$ as:
\begin{equation}
    \mathcal{L} = \mathcal{L}^{BPR} + \lambda_1 \mathcal{L}^{CL} + \lambda_2 ||\Theta||^2_2
\end{equation} 
where the  hyper-parameter $\lambda_1$ controls the contrastive loss; $\lambda_2$, is the regularization weight and $\Theta$ denotes the model's trainable parameters.
\section{Experiments} 
\label{sec-exp}

In this section, we conduct extensive experiments on benchmark datasets to validate the effectiveness of \rcn{} and investigate the importance of its components.

\subsection{Experimental Setup}
\label{sec-exp-setup}

\subsubsection{Datasets.} 

\begin{table}[t!]
    \caption{Statistics of Youshu, NetEase and iFashion datasets.}
    \label{tab:data}
    \centering
    \resizebox{0.85\columnwidth}{!}{
    \begin{tabular}{cccccccc}
        \toprule
        \textbf{Dataset} & $|\mathcal{U}|$ & $|\mathcal{I}|$ & $|\mathcal{B}|$ & $|\mathcal{E}_{\mathcal{UI}}|$ & $|\mathcal{E}_{\mathcal{UB}}|$ & \#Avg.I/B & $c-score$ \\
        \midrule
        Youshu           & 8,039   & 32,770  & 4,771     & 138,515     & 51,377        & 37.03   & 0.0812  \\
        NetEase          & 18,528  & 123,628 & 22,864    & 1,128,065   & 302,303       & 77.80   & 0.0599  \\
        iFashion         & 53,897  & 42,563  & 27,694    & 2,290,645   & 1,679,708     & 3.86 &  0.2917    \\
        \bottomrule
    \end{tabular}
}

\end{table}



According to popular studies \cite{chang2020bundle,ma2022crosscbr,ren2023distillation}, we adopt three benchmark datasets for the bundle recommendation evaluation, including: 
The \textit{Youshu} \footnote{https://www.yousuu.com/} dataset consists of bundles formed by lists of books that is purchased in each user session; 
the \textit{NetEase} \footnote{https://music.163.com/} dataset captures user-generated song lists as bundles; 
the \textit{iFashion} \cite{chen2019pog} dataset establishes outfits composed of clothes and accessories as bundles. 
The statistics of these datasets are shown in Table \ref{tab:data}. 
Specially, $c-score$ metric, according to \cite{ren2023distillation}, present the high consistency between user-bundle and user-item collaborative relations on iFashion but limited on Youshu and NetEase.

\subsubsection{Evaluation Metrics.}

Recall ($R@K$) and Normalized Discounted Cumulative Gain ($N@K$) are two commonly employed metrics for the performance of method in bundle recommendation task. 
$R@K$ measures the proportion of test bundles within the top-$K$ ranking list. 
$N@K$ manifests normalized discounted cumulative gain scores aimed at obtaining relevant items at higher positions on the ranking list.
Both metrics are employed with $K \in \{10, 20\}$ for performance validation. 
The smaller the top-$K$, the more clearly it demonstrates the model's practical recommendation performance.
Average performances in $5$ runs with various random initialization are reported. Comparisons are evaluated by two-tailed paired-sample Student’s t-test with $p-$value of $0.05$.

\subsubsection{Baselines.}

We compare \rcn{} to the three groups of state-of-the-art models:

\begin{itemize}
    \item \textbf{Traditional Bundle Recommendation} $(T)$: BPR \cite{rendle2012bpr}, LightGCN \cite{he2020lightgcn}, and DAM \cite{chen2019matching}. 
    \item \textbf{Multi-view Learning Bundle Recommendation} $(M)$: CrossCBR \cite{ma2022crosscbr}, BGCN \cite{chang2020bundle}, MIDGN \cite{zhao2022multi}, BundleGT \cite{wei2023strategy}, and EBRec \cite{du2023enhancing}. 
    \item \textbf{Distillation Bundle Recommendation} $(D)$: DGMAE \cite{ren2023distillation}.
\end{itemize}

For BPR and LightGCN, we simply consider bundles at the item level similar to previous reputable studies \cite{chen2019matching,du2023enhancing,sun2022revisiting}.
For MIDGN, embedding each user or bundle into $k$ chunks within different feature spaces presents challenges for optimizing performance on large datasets like iFashion, particularly when computational resources are limited.
For DGMAE, obtaining complete materials from the authors to ensure absolute reproducibility still faces some difficulties. 
To facilitate a fair comparison, the available results have been compiled from relevant reputable papers.

\subsubsection{Implementation details.} 

\rcn{} is implemented using PyTorch on NVIDIA P100 and T4 GPUs, with evaluated datasets in Table~\ref{tab:data}. 
Baseline methods are evaluated with the same below settings based on source code and results in reputable works \cite{chang2020bundle,ma2022crosscbr,wei2023strategy}.
The initial embedding size is configured as $64$. 
The learning rate is set as $1e-3$. 
We adopt Xavier initialization and Adam optimizer for trainable parameters. 
We tune the regularization weight $\lambda_2 \in \{1e-6, 4e-6, 1e-5, 4e-5, 1e-4, 4e-4\}$. The $\alpha, \gamma, \mu, \beta, \lambda_1 \in \left[0;1\right]$ are tuned by grid search.

\begin{table*}[t!]
    \caption{Performance comparison to baselines. $\star$ denotes values in the previous research paper, while others are in our experiments. The best performances is in \textbf{bold} and the second best values is \underline{underlined}. \textit{Imp} shows the relative improvements of \rcn{} over the second-best model. $\dagger$ indicates statistically significant improvements. ``$-$'' presents missing values due to the unavailability or run-time problems of the source codes. $(T)$, $(M)$, $(D)$ respectively represent the best results of each baseline group.
    }
    \label{tab:main-results}
    \centering
    \resizebox{1.0\textwidth}{!}{
        \begin{tabular}{l|cccc|cccc|cccc}
            \toprule
             Dataset&  
            \multicolumn{4}{c|}{\textbf{Youshu}} &  \multicolumn{4}{c|}{\textbf{NetEase}}&  \multicolumn{4}{c}{\textbf{iFashion}}\\
            \midrule
            Metric &  \textit{R@10}&\multicolumn{1}{c}{\textit{R@20}}  & \textit{N@10}& \multicolumn{1}{c|}{\textit{N@20}}      &  \textit{R@10}&\multicolumn{1}{c}{\textit{R@20}}  & \textit{N@10}& \multicolumn{1}{c|}{\textit{N@20}}      &  \textit{R@10}&\multicolumn{1}{c}{\textit{R@20}}  & \textit{N@10}& \multicolumn{1}{c}{\textit{N@20}}      \\ 
            \midrule
            BPR&  $15.03$&$21.49$&$10.11$& $11.96$&  $0.12$&$0.22$&$0.08$& $0.11$&  $0.46$&$0.81$&$0.41$& $0.55$\\
            LightGCN&  $16.93$&$24.34$&$12.22$& $14.35$&  $3.24$&$5.42$&$2.16$& $2.84$&  $3.25$&$5.33$&$2.92$& $3.78$\\
             DAM&  $11.03$&$20.82$&$8.73$& $11.98$&  $2.65$&$4.11$&$1.43$& $2.10$&  $3.17$&$6.29$&$2.66$& $4.50$\\
             \midrule
             BGCN& $15.55$ & $23.47^{\star}$ & $10.99$& $13.45^{\star}$& $2.88$ & $4.91^{\star}$ & $1.88$& $2.58^{\star}$ &  $4.83$ & $7.33^{\star}$ & $4.34$ & $5.31^{\star}$ \\
             BundleGT& \underline{$20.01$}& {$29.05^{\star}$}& {$14.16$}& {$17.37^{\star}$} & {$5.35$}& \underline{$9.03^{\star}$} & {$3.51$}& {$4.78^{\star}$}& {$8.16$}& {$12.63^{\star}$}& {$7.65$}& {$9.69^{\star}$}\\
             MIDGN & $18.87$ & $26.82^{\star}$ & $12.91$ & $15.27^{\star}$ &  $3.74$ & $6.78^{\star}$ & $2.46$ & $3.43^{\star}$ &  $-$ & $6.22^{\star}$ & $-$ & $4.41^{\star}$\\
             EBRec& $19.62$& $28.50$& $\underline{14.27}$& $16.80$&  $\underline{5.49}$& $8.83$& $\underline{3.69}$& $4.76$&  \underline{$9.01$}& \underline{$13.22$}& $\underline{8.57}$& \underline{$10.28$}\\
             CrossCBR&  $19.61$ & $28.13^{\star}$ & $14.21$ & $16.68^{\star}$ &  $5.11$ & $8.42^{\star}$ & $3.44$ & $4.57^{\star}$ &  $7.72$ & $11.73^{\star}$ & $7.31$ & $8.95^{\star}$ \\
             \midrule
             DGMAE& $-$& $\textbf{32.45}^{\star}$& $-$& $\textbf{19.18}^{\star}$&  $-$& $\textbf{9.45}^{\star}$& $-$& $\textbf{5.12}^{\star}$&  $-$& $12.62^{\star}$& $-$& $9.59^{\star}$\\
             \midrule
             \textbf{\rcn{}} & $\textbf{20.85}^{\dagger}$ &$\underline{29.78}$& $\textbf{14.65}^{\dagger}$& $\underline{17.62}$ & $\textbf{5.55}$ &$8.99$&$\textbf{3.73}$& $\underline{4.79}$& $\textbf{9.80}^{\dagger}$&$\textbf{14.31}^{\dagger}$&   $\textbf{9.39}^{\dagger}$& $\textbf{11.23}^{\dagger}$\\ 
             \midrule
            $\% \uparrow$ $(T)$
            & $23.15$& $22.35$& $19.88$& $22.78$& $71.29$& $65.86$& $72.68$& $68.66$& $301$& $227$& $321$&$250$\\
            $\% \uparrow$ $(M)$
            & $4.20$& $2.51$& $2.66$& $1.44$& $1.72$& $0.44$& $1.09$& $0.21$& $8.77$& $8.25$& $9.57$&$9.24$\\
            $\% \uparrow$ $(D)$
            & $-$& $-8.22$& $-$& $-8.13$& $-$& $-4.87$& $-$& $-6.44$& $-$& $13.39$& $-$&$17.10$\\
            \bottomrule
        \end{tabular}
    }
    
\end{table*}

\subsection{Performance Comparison to Baselines}
\label{subsec:comparision_baseline}

Table~\ref{tab:main-results} illustrates performance comparison between \rcn{} and related baselines in term of $R@K, N@K$ metrics with $K \in \{10, 20\}$. Generally, the recommendation methods with multi-view learning consistently outperforms the traditional techniques for all metrics. This observation implies the efficiency of the decomposition and modeling of multiple views for the bundle recommendation task. 

Among the multi-view learning models, \rcn{} achieves the best performance in the benchmark datasets. 
Compared to applied contrastive learning methods (e.g., MIDGN, BGCN, CrossCBR), \rcn{} represents persistent enhancements, which verify the effectiveness of assembling discrete and concrete contrastive learning in providing discriminative representation across different views.     
Compared to the second-best results (BundleGT, EBRec), \rcn{} produces statistically significant improvements in terms of \textit{$R@10$} and \textit{$N@10$}. 
Especially, for iFashion, the notable enhancements are $8.25\%$ to $9.57\%$ for both $K \in \{10, 20\}$. 
Compared to the DGMAE distillation framework, the performance of \rcn{} is $13\%$ and $17\%$ superior on $R@20$ and $N@20$ for iFashion, but weak for Youshu and NetEase.
They affirm the effectiveness of exploiting item causation-sensitive relationships in small-sized bundles, in which bundle items are more coherent. 

For the case of large-sized bundle datasets such as NetEase and Youshu, the superiority of \rcn{} could be negatively affected due to noisy connections.
In real-world scenarios, users tend to buy bundles less frequently compared to individual items  due to the sensitve prices. To effectively promote user consumption and business marketing strategy, the constructed bundles need to be oriented towards specific and explicit intentions \cite{wei2023strategy,zhu2014bundle,sun2022revisiting}. Therefore, advanced approaches that exploit the interaction between items to model the bundle's targets (e.g., BundleGT, EBRec, \rcn{}) are clearly more effective on iFashion compared to the other architectures. 
On the other hand, Youshu and Netease data are solely built from large pre-defined item sets via user session and the limited consistency between the U-B and U-I collaborative relations ($c-score$). 

\begin{figure}[t!]
    \centering
    \includegraphics[width=0.9\columnwidth]{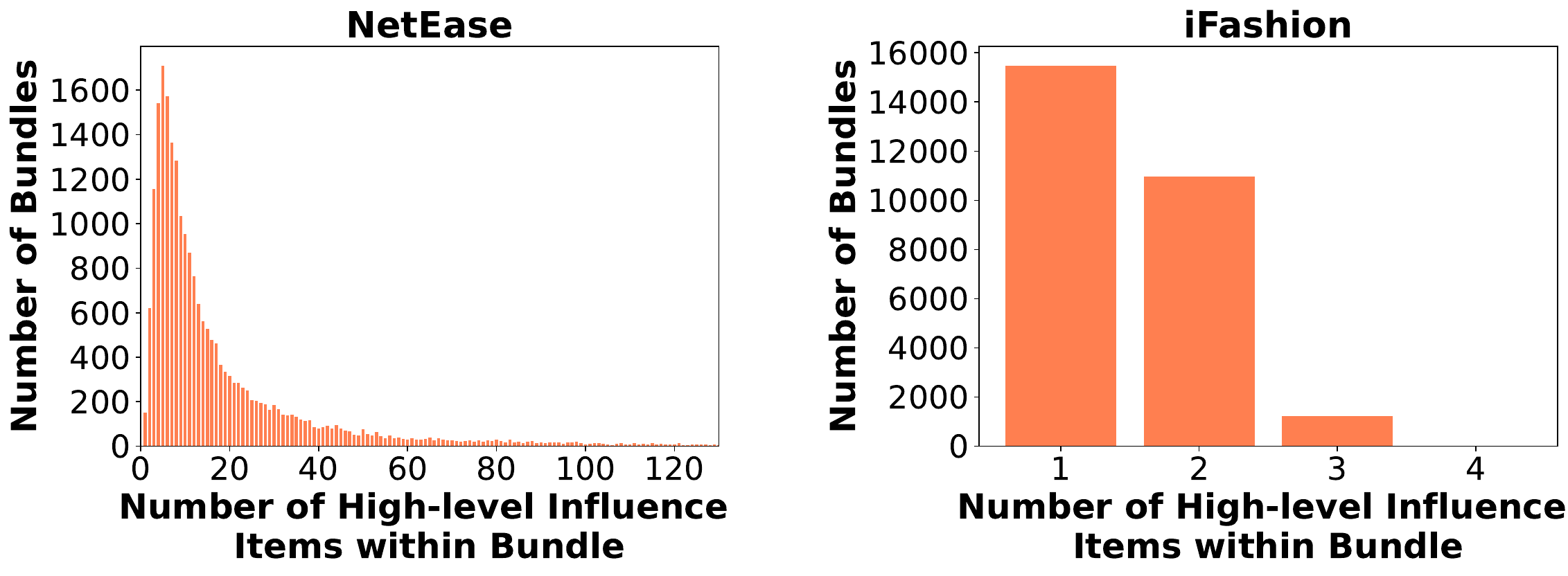}
    \caption{Statistical distribution of High-level Influence Items within Bundles.}
    \label{fig:high_influence_items}
\end{figure}

Moreover, we investigate the distribution of high-level influence items within bundles, determined by their popularity surpassing the average popularity of the items in the same bundle. 
According to Figure~\ref{fig:high_influence_items}, the distribution on iFashion indicates that each bundle's characteristics are mainly shaped by one or two key components.
Conversely, many items within the same bundle demonstrate high-level influence in NetEase dataset featuring large-sized bundles (similar to Youshu).
This phenomenon causes noise and hinders the identification of anchor items necessary for transparent bundle modeling.
DGMAE hardly models bundle-item affiliations in depth but only synthesizes bundle representations based on the average pooling of item representations \cite{ren2023distillation}.
Therefore, DGMAE's performance can be superior on Youshu and NetEase but are significantly limited on iFashion.
The above evidence points to a contrast with the practical intuition of bundle recommendation tasks. 
This statement demystifies the notable consideration of the reasonableness and quality of popular benchmark datasets for bundle recommendation, which is also mentioned recent studies \cite{sun2024revisiting,sun2022revisiting}.

\subsection{Ablation Study}
\label{sec-exp-model}
To investigate the impact of key components in \rcn{}, we conduct various ablation studies described below.

\begin{table}[t!]
    \caption{The performance of \rcn{} once omitting various sub-view  separately. The $\downarrow$ denotes the percentage decrease when each specific view is ablated from \rcn{}.}
    \label{tab:ablation}
    \centering
    \resizebox{0.85\columnwidth}{!}{
    \begin{tabular}{c|c|cccc|c}
        \toprule
        Dataset& \multicolumn{1}{c|}{Metric}& w/o RV& w/o SV& w/o UP& w/o BC& \textbf{\rcn{}}\\
        \midrule
        {\multirow{2}{*}{Youshu}}& 
            \textit{R@20}  & $26.54_{(\downarrow \, 10.84)}$& $26.06_{(\downarrow \, 12.46)}$& $29.38_{(\downarrow \, 1.32)}$& $27.88_{(\downarrow \, 6.36)}$&$\textbf{29.77}$\\& 
            \textit{N@20}    & $15.38_{(\downarrow \,12.69)}$& $15.03_{(\downarrow\, 14.70)}$& $16.85_{(\downarrow \,4.34)}$& $16.65_{(\downarrow \,5.47)}$&$\textbf{17.62}$\\ 
            \midrule
            {\multirow{2}{*}{NetEase}}& 
            \textit{R@20}  & $5.65_{(\downarrow \, 37.17)}$& $6.00_{(\downarrow \, 33.28)}$& $8.08_{(\downarrow \, 10.21)}$& $8.56_{(\downarrow \, 4.81)}$&$\textbf{8.99}$\\ & 
            \textit{N@20}    & $2.92_{(\downarrow \,39.00)}$& $3.11_{(\downarrow\, 35.03)}$& $4.39_{(\downarrow \,8.21)}$& $4.65_{(\downarrow\, 2.82)}$&$\textbf{4.79}$\\
            \midrule
            {\multirow{2}{*}{iFashion}}& 
            \textit{R@20}  & $5.36_{(\downarrow \,62.52)}$& $12.79_{(\downarrow \,10.61)}$& $14.20_{(\downarrow \, 0.74)}$& $11.48_{(\downarrow \, 19.79)}$&$\textbf{14.31}$\\ & 
            \textit{N@20}    & $3.77_{(\downarrow \,66.37)}$& $10.53_{(\downarrow\, 6.07)}$& $11.11_{(\downarrow 0.91)}$& $9.15 _{(\downarrow \,18.39)}$& $\textbf{11.21}$\\
        \bottomrule
    \end{tabular}
    }
    
\end{table}

\subsubsection{Effects of Learning from Different Views.}
To assess the impact of integrating latent representations from different views, learning Cohesive View (SV), Coherent View (RV), UP sub-view, and BC sub-view are selectively omitted and evaluated performance on the three datasets.
According to Table~\ref{tab:ablation}, the results indicate that learning representations from each view encodes valuable information to enhance prediction performance.
Excluding SV learning significantly diminishes \rcn{}'s performance on all datasets, demonstrating its ability to effectively leverage high-order collaborative signals between users and bundles. 
Besides, the performance drops remarkably once RV-view learning is omitted. It demystifies the effectiveness of learning causation-sensitive relationships between items to enhance latent representations. 
UP sub-view representations exhibit a marginal decrease of approximately $1\%$ when omitted on iFashion, contributing the least valuable information. 
Conversely, the performance of UP sub-view significantly affected on NetEase with about $10\%$.
The results on iFashion without the BC sub-view highlight the explicit bundle construction more than the results on Youshu and NetEase. 
This demonstration also contributes to clarifying the observation in Section~\ref{subsec:comparision_baseline}.
In summary, integrating latent representations through multi-view learning ensures both cohesion and coherence in bundle construction and user preferences, effectively improving performance.

\subsubsection{Effect of Contrastive Learning Module}

As depicted in Table~\ref{tab:contrastive}, we examine the effectiveness of contrastive learning module through experiments involving the removal of the entire module (\textit{w/o CL}), discrete contrastive learning (\textit{w/o DC}), and concrete contrastive learning (\textit{w/o CC}). 
The results across the three datasets simultaneously emphasize the pivotal role of modeling multi-view information integrated with contrastive learning.
Notably, the substantial impact of discrete contrastive learning demonstrates its effectiveness in aligning the representations of seperate views. 
This implies that each view can extract cooperative information from the other, leading to mutual enhancement.
The addition of concrete contrastive learning further confirms that the contrastive learning module can enhance the discrimination of each user/bundle in the fused representations.

\begin{table}[t]
    \caption{The impact of different contrastive learning modules on the \rcn{} performance. The $\downarrow$ denotes the ratio decrease when ablating each corresponding module.}
    \label{tab:contrastive}
    \centering
    \resizebox{0.7\columnwidth}{!}{
    \begin{tabular}{c|c|ccc|c}
        \toprule
        Dataset& \multicolumn{1}{c|}{Metric}& w/o CL& w/o DC& w/o CC& \textbf{\rcn{}}\\
        \midrule
        {\multirow{2}{*}{Youshu}}& 
            \textit{R@20}          & $27.49_{(\downarrow \,7.64)}$& $25.50 _{(\downarrow \,14.33)}$& $29.16_{(\downarrow \,2.06)}$&$\textbf{29.77}$\\& 
            \textit{N@20}            & $15.97_{(\downarrow\,9.34)}$& $15.03_{(\downarrow \,14.70)}$& $17.03_{(\downarrow\,3.34)}$&$\textbf{17.62}$\\
            \midrule
            {\multirow{2}{*}{NetEase}}& 
            \textit{R@20}          & $5.78_{(\downarrow \,35.71)}$& $6.55_{(\downarrow \,27.16)}$& $8.72_{(\downarrow \,3.00)}$&$\textbf{8.99}$\\& 
            \textit{N@20}            & $3.05_{(\downarrow \,36.30)}$& $3.61_{(\downarrow \,24.52)}$& $4.67_{(\downarrow \,2.54)}$&$\textbf{4.79}$\\
            \midrule
            {\multirow{2}{*}{iFashion}}& 
            \textit{R@20}              & $6.39_{(\downarrow \, 55.35)}$& $7.25_{(\downarrow \,49.35)}$& $13.46_{(\downarrow\,5.91)}$&$\textbf{14.31}$\\& 
            \textit{N@20}                & $4.55 _{(\downarrow\,59.42)}$& $5.33_{(\downarrow \,52.51)}$& $10.28_{(\downarrow \,8.32)}$& $\textbf{11.21}$\\
        \bottomrule
        \end{tabular}
        }

\end{table}

\subsubsection{Importance of Asymmetric Causation Matrix}

\begin{figure}[t!]
    \centering
    \includegraphics[width=1.0\columnwidth]{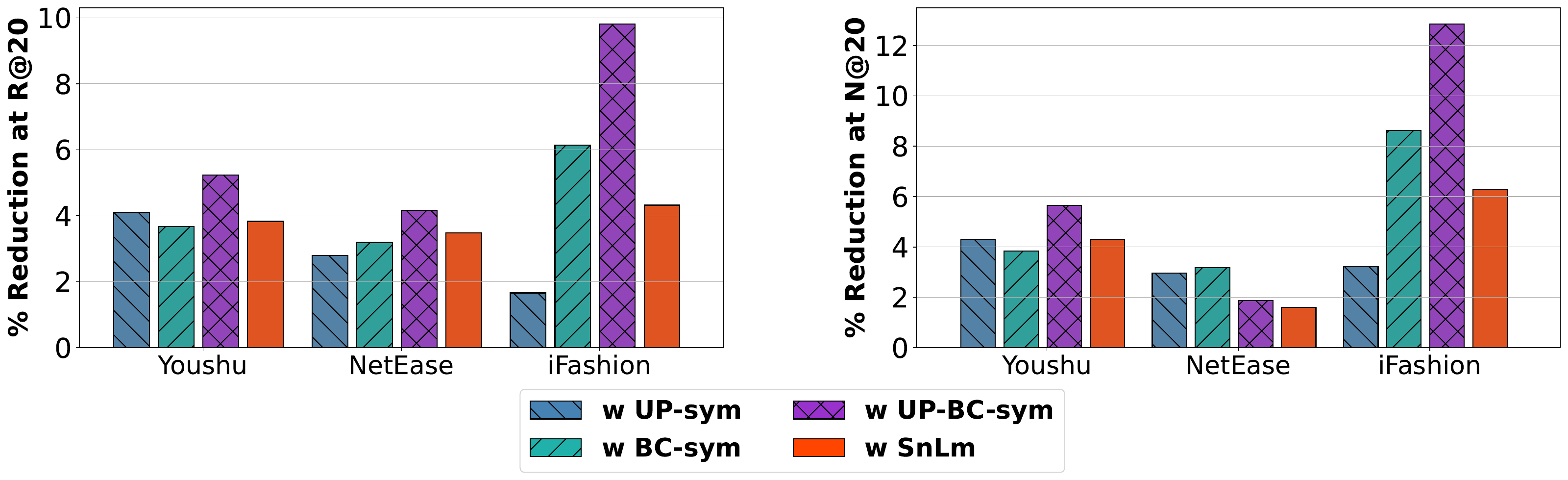}
    \caption{Importance of asymmetric causation matrix $A$ on the performance of \rcn{}.}
    \label{fig:asym-chart}
\end{figure}

The causation matrix $A$ learned by the MPCNet module enhances item-level representation ${t_i^f}$ for bundle recommendation. 
In order to assess its significance, we compare it against alternatives: i) the original symmetric co-occurrence matrix from UP sub-view (\textit{w UP-sym}) or BC sub-view (\textit{w BC-sym}); ii) both original symmetric co-occurrence matrices from UP and BC sub-views (\textit{w UP-BC-sym}); iii) normalized Laplacian symmetric correlation matrix as in \cite{le2019correlation} (\textit{w SnLm}).
Figure~\ref{fig:asym-chart} shows a notable performance decrease of \rcn{} without asymmetric matrix-enhanced item representations, instead simply employing symmetric item-item matrices.
This demonstration validates the significance of asymmetric relationships in real-world scenarios, affirming the hypothesis discussed in Section~\ref{sec:intro}. 


\subsubsection{Effects of key hyper-parameters}

\begin{figure}[t!]
    \centering
    \includegraphics[width=0.86\columnwidth]{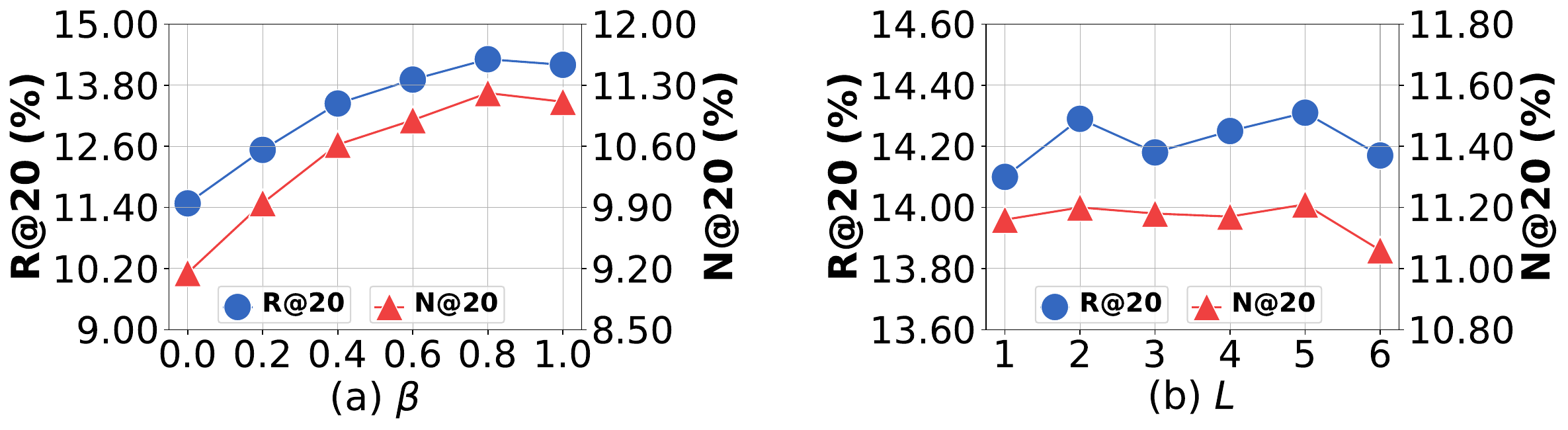}
    \caption{The impact of key hyper-parameters on the performance of \rcn{} for iFashion. Figures (a) and (b) show the model performance calculated by $R@20$ and $N@20$ over the respective value ranges of $\beta$ and $L$.}
    \label{fig:hyper-param}
\end{figure}

In Figure \ref{fig:hyper-param}, the impact of key hyperparameters, $\beta$ and $L$, on \rcn{}'s performance for iFashion is depicted. 
The parameter $\beta$ regulates the contribution of enhanced item representations from the BC sub-view and UP sub-view. 
As $\beta$ increases, the performance follows an ascending trend until reaching its peak at $\beta = 0.8$, underscoring the crucial role of the BC sub-view in refining item representations. 
Regarding $L$, the performance exhibits minor fluctuations within the range of 1 to 5 prospects before a decline is observed with $L = 6$. 
This decline is attributed to noise problems that increase simultaneously with latent features of high-order neighbors. 
After tuning, the optimal settings for iFashion are determined to be $\beta = 0.8$ and $L = 5$. 
Similar tuning processes for Youshu and NetEase yield comparable observations.

\subsection{Qualitative Showcase}

\begin{figure}[t!]
    \centering
    \includegraphics[width=1.0\columnwidth]{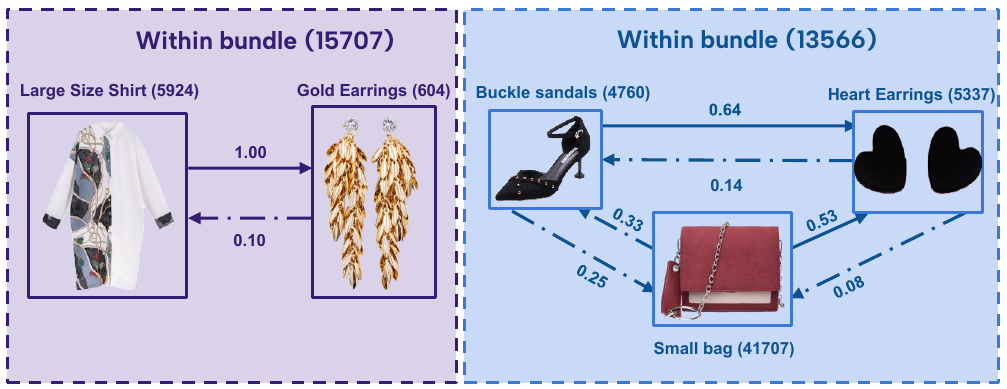}
    \caption{The illustration of real cases in iFashion. Solid arrows indicate highly affected causation with asymmetric weight $ \ge 0.5$ whilst dash-dotted ones denote low effects.}
    \label{fig:showcase_example}
\end{figure}

Some qualitative examples in the iFashion dataset is illustrated in Figure~\ref{fig:showcase_example} due to the coherence and specificity of interconnected items, whereas Youshu and NetEase only provide the IDs of objects.
The asymmetric weights indicate the difference effect among items, resulting in determining the explicit target of bundle features.
Thereby, within each bundle, certain anchor items can be identified, and their influence on complementary items varies depending on causation signals.
For instance, considering the bundle $15707$ containing a \textit{shirt} and \textit{earrings}. The weights reveal that the \textit{shirt} serves as the anchor item, as evidenced by its influence on the \textit{earring} with a weight of $1.0$, while the impact of the \textit{earring} on the \textit{shirt} is only $0.1$ in asymmetric weight. Likewise, the bundle $13566$ shows reasonable causation weights from \textit{buckle sandals} and a \textit{small bag} to \textit{heart earrings}. 
These evidence can be readily understood through the decision-making psychology of female customers.
This emphasizes the ability to model key components within bundles via asymmetric relationships that adapt to user engagements. 
Therefore, it validates our hypothesis in leveraging item-level causation effects for bundle recommendation.

\section{Conclusion}
\label{sec-con}
This study design a novel approach to address bundle recommendation by emphasizing the significance of asymmetric relationships between items. 
Our proposed model \rcn{} leverages item-level causation-enhanced multi-view learning, showcasing significant improvements and notable points, as evidenced by extensive analytics conducted on three benchmark datasets.
The architecture of \rcn{} lies in learning two distinct views: the Coherent View, employing the Multi-Prospect Causation Network for causation-enhanced representation learning; and the Cohesive View, utilizing high-order collaborative signals of user-bundle interaction to aggregate comprehensive representations.
In addition, \rcn{} simultaneously integrates concrete and discrete contrastive learning, enhancing the consistency and self-discrimination of representations from both views. 
Our work not only validates the efficacy of \rcn{} but also emphasizes the importance of considering asymmetric item relationships for bundle-related tasks. 


\begin{credits}
\subsubsection{\ackname} 
This research was funded by the Vingroup Innovation Foundation under the project code VINIF.2022.DA00087 and the Master \& PhD scholarship program code VINIF.2023.ThS.114. 


\end{credits}

%
%
%

\end{document}